# Low Temperature Magnetic Domain Patterns in MnAs Films Grown on GaAs(001)

M. Cheon, S. Hegde, S. Wang, M.M. Bishara, G.B. Kim, and H. Luo[a]
*Department of Physics, University at Buffalo, The State University of New York, Buffalo, NY, 14260*

Magnetic properties of MnAs were studied as a function of temperature with a superconducting quantum interference device (5 K to 350 K), and atomic force microscopy/magnetic force microscopy (20 K to 360 K). Structural and magnetic properties of MnAs depend on film thickness both near and far below the Curie temperature. In samples with coexisting ferromagnetic α–MnAs and paramagnetic β–MnAs, the domain structures are affected by the distribution of the two phases. The magnetic domain structures below the temperature range of this mixed phase resemble that of a single domain structure with uniform magnetization along the easy axis, except there are regions elongated along the easy axis embedded where the magnetization is along the second easy axis, i.e., normal to the films. The shape of those regions and their temperature dependence are also related to the MnAs layer thickness.

Ferromagnetic properties of MnAs is of great interest in the context of spintronics because its Curie temperature is above room temperature (~320 K) and its structural compatibility with GaAs(001).[1,2] The growth, structural, magnetic and electronic properties have been investigated for spintronic applications.[3-6] Its potential as a component of spintronic devices depends critically on its magnetic domain patterns.

Structural and magnetic domain properties of MnAs have been extensively studied near room temperature.[7-10] When MnAs is grown on GaAs(001), both hexagonal ferromagnetic α–MnAs and orthorhombic paramagnetic β–MnAs can be present, depending on the thickness and temperature. Although the two phases follow a first order transition in bulk MnAs, it occurs over a temperature range, typically a few tens of degrees Kelvin, close to the Curie temperature. We will refer to this coexistence of the two phases as the mixed phase. In relatively thick MnAs epilayers, *e.g.*, thickness ~ 180 nm, α–MnAs and β–MnAs form quasiperiodic stripe patterns at room temperature, with stripes perpendicular to the easy axis.[11]

The magnetic domain patterns of MnAs samples have been studied for the mixed phase, from 27 °C to 41 °C, revealing the connection between structural features and magnetic properties.[10] The domain patterns in samples with MnAs thickness around 180 nm are characterized by small domains distributed within the α–MnAs stripes with magnetization along the easy axis, perpendicular to the stripes. In thinner sample (thickness ~ 60 nm), α–MnAs were found to be single domain within the stripes.[9] The observed magnetic domain patterns in the mixed phase is not desirable for use as a spin injector. One important question is what the domain patterns are like below the temperature range of the mixed phase.

The structural and magnetic properties of α–MnAs and β–MnAs phases affect each other near the Curie temperature. Thus the magnetic domain patterns are quite complex near the phase transition. Because both the concentration and the distribution of α–MnAs and β–MnAs phases depend on film thickness, the difference in magnetic properties of samples having different thicknesses will shed light on their dependence on the structural features. In this work, we studied samples of different thicknesses over a wide temperature range, focusing on magnetic domain patterns at temperatures below the temperature range of the mixed phase.

The MnAs epilayers used in this work were grown in a Riber 32P molecular beam epitaxy system on GaAs(001) substrates, at a substrate temperature of 250°C. The orientation of the MnAs was determined to be MnAs($\bar{1}$100)||GaAs(001) and MnAs[0001]||GaAs[1$\bar{1}$0]. The MnAs thicknesses for the samples used are 10 nm, 40 nm, 80 nm and 400 nm, which will be referred to as Samples 1, 2, 3, and 4, respectively. The thicknesses were measured with transmission electron microscopy (TEM) for Sample 1, because of its small thickness and scanning electron microscopy (SEM) for Samples 2-4.

The magnetic properties were studied, as a function of temperature, with a superconducting quantum interference device (SQUID) and an Omicron variable temperature scanning probe microscope operated in the atomic force microscopy (AFM) and magnetic force microscopy (MFM) modes. Because of the temperature range of the ferromagnetic phase in the samples studied, the experiments were performed at temperatures from 25 K to 360 K. For surface structures, contact mode AFM was used. Because of the strong magnetic signal, images taken with noncontact mode AFM contain a great deal of magnetic features.

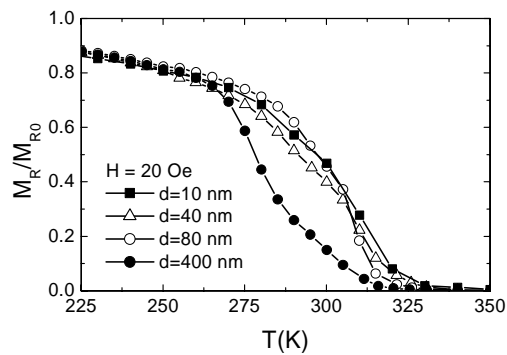

Figure 1. The temperature dependence of remanent magnetization in as-grown MnAs epilayers of different thicknesses.

The temperature dependence of remanent magnetization ($M_R$) of the four samples is shown in Fig. 1, normalized to their respective values at 5 K ($M_{R0}$) for comparison. The behavior of the remanent magnetization in Samples 1-3 is similar to each other. Sample 4 showed a more rapid decrease above 270 K. The AFM/MFM measurements showed an onset of α–MnAs to β–MnAs phase transition around 260 K with rising temperature in Sample 4, far below the temperature reported so far (for thinner samples). The rapid reduction of the remanent magnetization is a combination of the reduced magnetization of α–MnAs approaching the Curie temperature and the decreasing volume of α–MnAs.

Among the samples studied, it was observed that the stripe width increases with increasing layer thickness. The average periods for the stripes in Samples 1-4 are approximately 230, 290, 570, and 1200 nm, respectively, consistent with theoretical results.[7] A comparison between the period in Sample 4 and that reported for a 180 nm thick sample in Ref. 7 suggests that the increase of the periods reaches a maximum when the thickness is over 180 nm. This clearly affects magnetic domain patterns, because the larger periods also correspond to more complex domain structures within the α–MnAs stripes.

We will first focus on Sample 4. At room temperature, its magnetic domain pattern is characterized by small domains distributed within the α–MnAs stripes, with magnetization pointing in both directions along the easy axis, in agreement with reported patterns for samples over 100 nm.[10,11] When the temperature is reduced, the α–MnAs stripes widen, and the β–MnAs stripes correspondingly narrow. Although the bulk structural phase transition (first order phase transition) between the two phases occurs at 313 K,[12] previous work showed that in thin films grown on GaAs(001) the mixed phase extends over a temperature range of approximately 20K, attributed to the strain in the epilayers.[6,7,13] In our experiments, it was found that this temperature range can be significantly greater for thicker samples.

The temperature dependence study of Sample 4 reveals both the magnetic domain structures at low temperatures and the source of its different temperature dependence of remanent magnetization compared to the other samples. For all MFM studies, the samples were magnetized along the easy axis and then cooled with zero field. Three MFM images of the same area of Sample 4 are shown in Fig. 2, taken at 20 K, 262 K and 295 K. It should be pointed out that the measurements were taken with rising temperature because there is a temperature hysteresis[7], *i.e.*, the composition of the mixed phase depends on the direction of the temperature sweep. The pattern at 295 K is similar to those presented in Ref. 11 for a 180 nm thick MnAs epilayer. When the temperature is below 260 K, the patterns consist of a continuous domain throughout the sample (the background) magnetized in the direction of initial applied field along the easy axis, with regions (both dark and bright areas) distributed randomly. These areas have similar characteristics, elongated along the easy axis (and also the direction of the overall magnetization). This is an energetically favorable configuration because it minimizes field divergence (or the amount of magnetic charge) along the domain walls. There is no visible change of the pattern at temperatures from 20 K to 260 K.

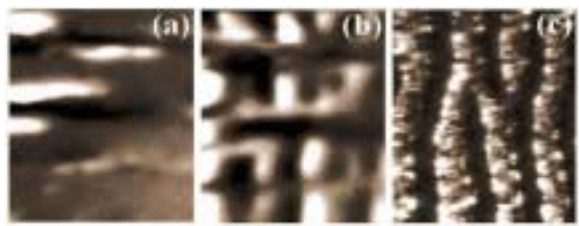

Figure 2. MFM images of the same area of Sample 4 taken at 20 K (a), 260 K (b), and 295 K (c). Scan size is 6 $\mu$m × 7 $\mu$m, and easy axis is horizontal.

The measurements were carried out with the magnetization of the MFM tip pointing into the sample surface. To understand the magnetic domains at low temperatures, numerical simulations were carried out with parameters used in the experiment, assuming that the magnetic field from the tip does not affect the magnetic properties of the sample. In particular, the simulation was done for the image shown in Fig. 2(a). In our MFM measurements, MFM contrast is the result of a change in cantilever oscillation frequency, which is proportional to the derivative of the normal component of the total tip-sample interaction force,[14]

$$F_z(r) = \int_{tip} (M_z(r) \cdot \frac{\partial^2}{\partial z^2} B_z(r-r') d^3r) ,$$

where $z$ is the out of sample plane direction, $M_z(r)$ the magnetic moment of the tip in $z$-direction (as used in the experiments) and $B_z$ the $z$-component of the magnetic field of the sample. The directions of magnetization for various regions used in the simulation are indicated in Fig. 3(b), after an extensive elimination process of other possibilities.

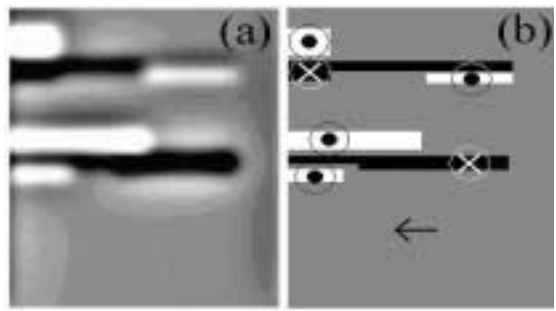

Figure 3. Numerical simulation of the image at 20 K, with bright areas and dark areas having spins aligned into and out of the sample surface, respectively.

The good agreement between Figs. 2(a) and 3(a) shows areas magnetized both into and out of the plane. This may be due to the fact that the normal of the sample surface is the second easy axis.[10]

When the temperature is raised to 262 K, the MFM image shows the onset of features related to the stripe patterns of the mixed phase that will eventually dominate at 295 K. Signs of vertical stripes (perpendicular to the easy axis) start to emerge at this temperature, while the low temperature pattern remain visible. The randomness of the details of the quasiperiodic pattern of α–MnAs and β–MnAs phrases is reproducible after the sample is cooled to the single α–MnAs phase and then warmed back to room temperature. It is an indication that the pattern is related to structural features, such as defects, that do not change significantly during the phase transition. Because the whole epilayer undergoes this transition, it is likely that the interface between the epilayer and the substrate is responsible for the details of the stripe patterns. Meanwhile, no correlation was found between the low temperature magnetic domain patterns and surface features.

A comparison between the temperature dependence of AFM/MFM and remanent magnetization of Sample 4 shown in Fig. 1 reveals that the rapid decrease of remanent magnetization around 265 K occurs at the same temperature when the paramagnetic magnetic β–MnAs stripes emerge. The combination of decreasing magnetization of the α–MnAs phase while approaching its Curie temperature and an earlier appearance of β–MnAs in Sample 4 leads to a faster down turn of its remanent magnetization.

As discussed earlier, the average period of the quasiperiodic pattern of the two phases depends on the film thickness. The smaller stripe width results in single domain α–MnAs stripes. The AFM/MFM measurements as a function of temperature showed that the transition from mixed phase to single α–MnAs phase occurs at a higher temperature for thinner samples, in agreement with the X-ray results reported in Ref. 7. Layer thickness also affects the magnetic domain patterns in the single α–MnAs phase at low temperatures. The low temperature elongated patterns seen in Sample 4 now become further elongated with decreasing layer thickness, appearing as long lines along the easy axis. As an example, an MFM image of Sample 2 at 284 K and an AFM image were shown in Fig. 4. An apparent decrease of the lengths of the lines with lowering temperature was observed. The lines disappeared when the temperature is lowered to 260 K, leaving a single domain structure, in contrast to the case of Sample 4 in which no significant change was observed in the single α–MnAs phase.

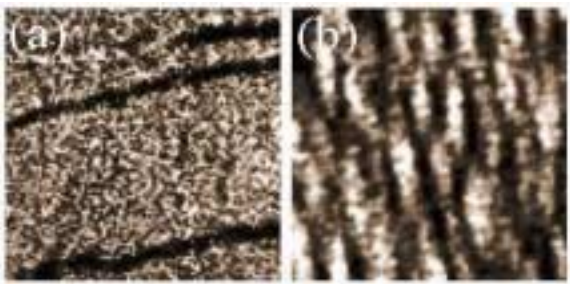

Figure 4. MFM(a) image of the Sample 2 taken at 284 K and AFM(b). Scan size is 1.5 $\mu$m × 1.5 $\mu$m and easy axis is horizontal.

In conclusion, we have investigated MnAs epilayers grown on GaAs(001). The results show that the magnetic domain patterns in the single α–MnAs phase below the Curie temperature have areas where the magnetization is not aligned along the easy axis. Instead, they are aligned perpendicular to the easy axis and the sample surfaces, *i.e.*, along the second easy axis. These areas are more thermally stable in thicker samples. Thinner samples (thickness less than 40 nm) become single domain once the temperature is well below the range of the mixed phase. The temperature range of the mixed phase in Sample 4 (from 262 K to 325 K) is much greater than the ranges reported earlier, which were all on substantially thinner samples.

This work was supported by NSF through ECS-0224206 and DARPA Spins Program through the Office of Naval Research under Grant N00014-00-1-0951.